\documentclass[prb,twocolumn,showpacs]{revtex4}
\usepackage{bm}
\usepackage{graphicx}

\begin{document}
\title{Reply to Comment on ``Spherical 2+p spin-glass model:
 an analytically solvable model with a glass-to-glass transition''}

\author{A. Crisanti$^\bullet$}
\email{andrea.crisanti@phys.uniroma1.it}

\author{L. Leuzzi$^{\bullet,\star}$} \email{luca.leuzzi@roma1.infn.it}
\affiliation{$^\bullet$ Dipartimento di Fisica, Universit\`a di Roma
``Sapienza'', P.le Aldo Moro 2, I-00185 Roma, Italy}
\affiliation{$^\star$ Statistical Mechanics and Complexity Center
(SMC), INFM - National Research Council (CNR), Italy}

\begin{abstract}
In his Comment, Krakoviack [Phys. Rev. B (2007)] finds that the phase
behavior of the $s+p$ spin-glass model is different from what proposed
by Crisanti and Leuzzi 
[Phys. Rev. B {\bf 73}, 014412 (2006)] if $s$ and $p$ are
larger than two and are separated well enough. He proposes a
trial picture, based on a one step replica symmetry breaking solution,
displaying a mode-coupling-like glass-to-glass transition line ending
in a $A_3$ singularity.  However, actually, the physics of these
systems changes when $p-s$ is large, the instability of which the one
step replica symmetry breaking glassy phase suffers turns out to be so
wide ranging that the whole scenario proposed by Krakoviack must be
seriously reconsidered.
\end{abstract} 


\maketitle

The model under consideration consists of $N$ spherical spins, i.e.,
 continuous variables $\sigma_i$ ranging from $-\infty$ to $\infty$,
 obeying the spherical constraint
\begin{equation}
\sum_{i=1}^N \sigma_i^2 = N, 
\end{equation}
and interacting via two different random quenched multi-body
interactions.  The Hamiltonian of the spherical $s+p$ spin glass model
is, then, defined by
\begin{equation}
\label{Ham}
{\cal H} =  \sum_{i_1<\ldots <i_s}J^{(s)}_{i_1\ldots i_s}
	   \sigma_{i_1}\cdots\sigma_{i_s}
           +\sum_{i_1<\ldots <i_p}J^{(p)}_{i_1\ldots i_p}
	   \sigma_{i_1}\cdots\sigma_{i_p}
\end{equation}
where $ J^{(t)}_{i_1 i_2..i_{t}}$, $t=s,p$, are uncorrelated zero mean 
random Gaussian variables of variance
\begin{equation}
\label{varVp}
   \overline{\left(J^{(t)}_{i_1 i_2..i_{t}}\right)^2} = 
    \frac{J_t^2 t!}{2N^{t-1}}, \qquad i_1 < \cdots < i_t
\end{equation}
The scaling with $N$ guarantees a correct thermodynamic limit.  As one
can see from Eq. (\ref{Ham}), we are considering the mean-field
approximation, in which each spin interacts with all other
spins. Notice that only distinct $s$-uples and $p$-uples are taken
into account in the Hamiltonian.

The properties of the spherical $s+p$  model strongly depend on
the values of $s$ and $p$: for $s=2$, $p=3$ the model reduces to the
usual spherical $p$-spin  model in a field\cite{CriSom92}
with a low temperature one step Replica Symmetry Breaking (RSB) phase
(i.e., the "mean-field glass" phase), while for $s=2$, $p\geq 4$ the
model possesses an additional Full RSB low-temperature
phase.\cite{Nieuwenhuizen95}
The model case under investigation
will be here the one with {\em both $s$ and $p$ larger than} $2$.

One of the interesting features of the model is that by defining the
auxiliary thermodynamic parameters $\mu_p=p\beta^2J_p^2/2$ a
straightforward connection can be made with the mode-coupling theory
(MCT). In the high temperature regime (i.e., in the fluid phase),
indeed, the dynamic equations of the model can be formally rewritten
as MCT equations in terms of the $\mu$'s and an exact mapping can be
set with mode coupling schematic theories $F_{s-1,p-1}$ with a scalar
kernel. In particular, the $F_{13}$ theory studied by G\"otze and
Sj\"ogren\cite{GoeSjo89} is dynamically equivalent to a $2+4$
spherical spin model.

\begin{figure}[t!]
\includegraphics[width=.45\textwidth]{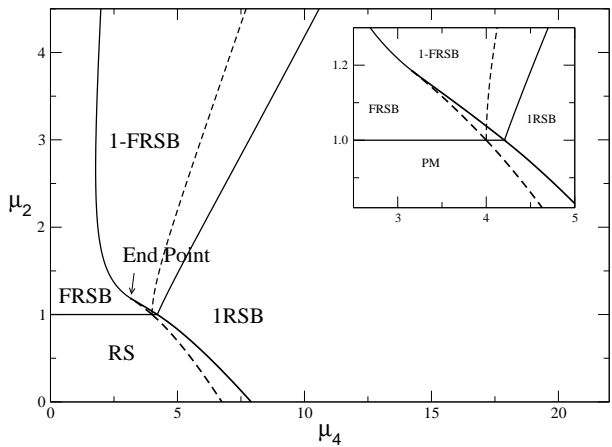}
\caption{Phase diagram of the spherical $2+4$ spin glass model.  RS:
replica symmetric (paramagnetic) phase; 1RSB: one-replica symmetry
breaking phase; FRSB: full replica symmetry breaking phase; 1-FRSB:
one-full replica symmetry breaking. The dashed lines refer to dynamic
transitions.  The continuous transition between the RS and the FRSB
phases and between the FRSB and 1-FRSB phases are the same for statics
and dynamics. Inset: close-up of the region around the ``end point''.
}
\label{fig:phdi.2+4}
\end{figure}

A partial analysis of the phase space of the $2+4$ model was carried
out in Ref. \onlinecite{CiuCri00} where, however, only the dynamical
stability of the 1RSB phase was considered leaving out a large part of
the phase space and, in particular, the question of the transition
between the 1RSB and the FRSB phases.  In that analysis, indications
for a glass-to-glass (G-G) transition line, called A$_2$ line in MCT,
ending in a A$_3$ point (namely, a cusp) was found. However, the whole
line was shown to remain in a region of the parameter space where the
1RSB phase was proven unstable.  The study of the dynamics and of the
statics of the generic spherical $2+p$ spin model has been completed
in further works by the authors,\cite{CriLeu04,CriLeu06,CriLeu06b},
especially showing that a different ``glassy'' phase arises (the ``1-Full''
RSB phase), different from the 1RSB one, as well as a Full RSB phase,
typical of systems with discrete spins (see, e.g.,
Refs. \onlinecite{SheKir75,Parisi80,Gardner85,Gross85}). In
Fig. \ref{fig:phdi.2+4} we reproduce the phase diagram of the $2+4$
model\cite{CriLeu06} to illustrate this.

If both $s$ and $p$ are larger than $2$, it was shown in
 Ref. \onlinecite{CriLeu06} that no 1-FRSB, nor FRSB phases occur and
 one has a smooth transition between a fluid/paramagnetic (PM) and a
 glassy phase (1RSB) (first dynamic, then static, lowering the
 temperature).  Krakoviack\cite{comment} pointed out in his Comment
 that this is true only as far as $p-s$ is not too large.  Indeed, he
 analyzes a parameter region of the spherical $s+p$-spin model not
 considered before, where the difference between $s$ and $p$ is large
 (in particular $s=3$, $p=16$) providing hints for a transition
 between two glassy phases both presumably described by a 1RSB
 solution.  A transition line of the type A$_2$, was drawn in the
 phase diagram terminating in a A$_3$ point.\cite{comment} In the RSB
 theory language, this corresponds to a point where a cusp between a
 physical and a non-physical branch of the constant $m=1$-line occurs,
 where $m$ is the value of the RSB parameter $x$ at which the overlap
 step function $q(x)$ discontinuously changes value.  This behavior is
 very reminiscent of what happens in the above-mentioned $2+p$
 spherical spin model.

\begin{figure}[t!]
\includegraphics[width=.47\textwidth]{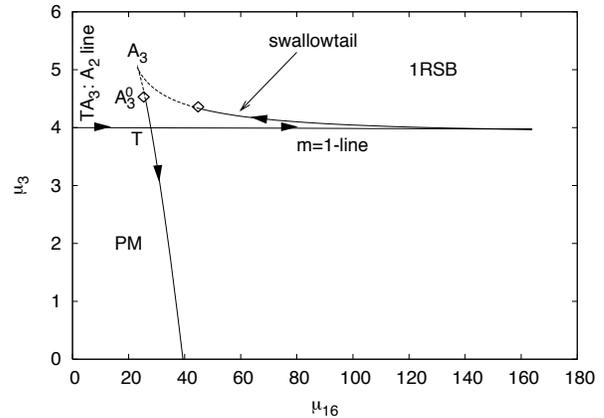}
\caption{Plot of the constant $m$ line at $m=1$,
 cf. Eqs. (\ref{f:m1p})-(\ref{f:m1s}), in the $\mu_{16}-\mu_{3}$ phase
 diagram. This is the dynamic transition line between the paramagnetic
 (PM) and the glassy 1RSB phases. The portion between the points T and
 A$_3$ is called A$_2$ line in MCT.  Following the arrows from
 $\mu_{16}=0$ the complexity first decreases below zero and then
 increases.  The dashed part of the curves correspond to points of the
 phase diagram for which the complexity is negative. }
\label{fig:3_16_a}
\end{figure}

The dynamic transition line (constant $m=1$) is plotted in
Fig. \ref{fig:3_16_a}, to be compared with Fig. 1 of
Ref. \onlinecite{comment}. It is determined by solving the equations
\begin{eqnarray}
\label{f:m1p}
\mu_p&=&\frac{(s-1)q_1-(s-2)}{(p-s)q_1^p-2(1-q_1)^2}
\\
\mu_s&=&-\frac{(p-1)q_1-(p-2)}{(p-s)q_1^s-2(1-q_1)^2}
\label{f:m1s}
\end{eqnarray}
For large enough $p-s$ the line displays, e.g., for $s=3, p=16$
 (Fig. \ref{fig:3_16_a}), the so-called swallowtail.

 Let us now consider in more detail this dynamic transition line.  In
  particular, we concentrate on the vertical part of the line above
  the crossing point: the A$_2$ line ${\overline {\rm TA}}_3$, cf.
  Fig. \ref{fig:3_16_a}. This is the first of the G-G lines proposed
  by Krakoviack (Fig. 1 of Ref. \onlinecite{comment}).  In Fig.
  \ref{fig:3_16_a} it is shown that, above a given $\mu_s$ value, the
  solutions along the line correspond to states of negative
  complexity, excluding, therefore, the A$_3$ point and reducing the
  length of the candidate G-G transition line (to an upper bound that
  we denote by A$_3^0$, where the complexity is zero).

\begin{figure}
\includegraphics[width=.47\textwidth]{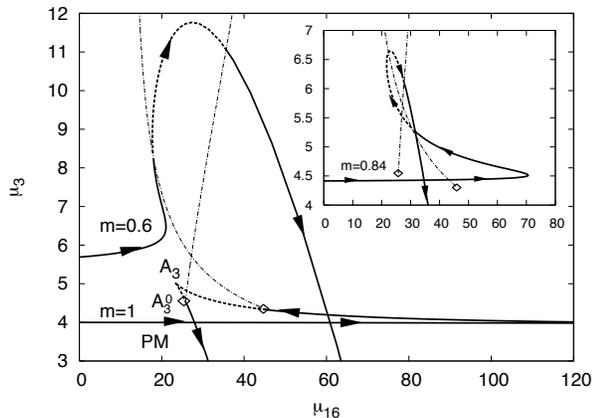}
\caption{Zero complexity curves and constant $m$-lines.  Three
$m$-lines are explicitly plotted as examples: $m=1$ (dynamic
transition), $m=0.6$ and $m=0.84$ (in the inset).  As a guide for the
eye we have put arrows: starting from $\mu_{16}=0$, $\Sigma$ decreases
from positive to negative values and, then, increases again to
positive values. With respect to the $\Sigma = 0$ curve, the side of
the phase diagram where the points of positive and negative complexity
are, depends on the value of $m$. As $m$ increases towards one, the
$m$-lines become more and more entangled.  The full curves represent
the parts of the $m$-lines whose points have $\Sigma>0$, the dashed
curves represent the points with $\Sigma<0$.  The two zero-complexity
curves (dashed-dotted lines) start from the $1$-line (empty diamonds).
}
\label{fig:3_16_c}
\end{figure}

  Analyzing the model further on, one can actually observe that, in
the model cases where the swallowtail and A$_2$ G-G line show up
(e.g., for $s=3$ it occurs when $p\geq 10$), in some region of the
phase diagram the hypothesized 1RSB solution turns out to be
inconsistent and must be, therefore, substituted by different, more
complicated {\em Ans\"atze}.  One sufficient signature of this
instability, by no means necessary, is, e.g., that the complexity of
the system becomes negative.  We show in Fig. \ref{fig:3_16_a} the
constant $m=1$-line, i.e., the dynamic PM/1RSB transition
line. Starting from $\mu_{16}=0$ (and following the arrows) the complexity
of the system first decreases from positive to negative and, afterwords,
increases along the vertical branch, becoming positive again.  Part of
the A$_2$ line thus corresponds to a system with negative complexity.
\\ \indent A serious danger exists that the instability might extend
 into the frozen region, heavily affecting the G-G line, similarly to
 what occurred in the case $s=2$.\cite{CriLeu04,CriLeu06} Deepening
 the analysis of the model one finds out that this is actually what
 happens.
\\ \indent To show how the instability occurs, we first begin with the
 1RSB solution, looking at its ``bugs''.  In Fig. \ref{fig:3_16_c} the
 loci of zero complexity are reported. These lines are constructed
 following the $m$-lines with $m\leq 1$ as we did above for the
 $1$-line.  For small $m$ the $m$-lines are single-valued in the
 ($\mu_{16}$, $\mu_3$) plane and the complexity $\Sigma$ computed
 along an $m$-line first becomes negative and then positive as
 $\mu_{16}$ increases.  Increasing $m$, the $m$-lines  start to be multivalued
 in some parameter region (see the curve for $m=0.6$ in
 Fig. \ref{fig:3_16_c}).  Moreover, as $m$ further increases to
 $m\lesssim 1$, the generic $m$-line crosses itself and, always starting
 from $\mu_{16}=0$, the point at which $\Sigma$ first decays to zero
 turns out to be on the right side with respect to the point where
 $\Sigma$ reaches zero from below (look, e.g., at the inset of
 Fig. \ref{fig:3_16_c}).  Spanning the plane with $m$-lines, two lines
 of loci of zero complexity system points are constructed, signaling
 that the 1RSB solution is not the physical one and must be rejected
 and substituted.

\begin{figure}[t!]
\includegraphics[width=.47\textwidth]{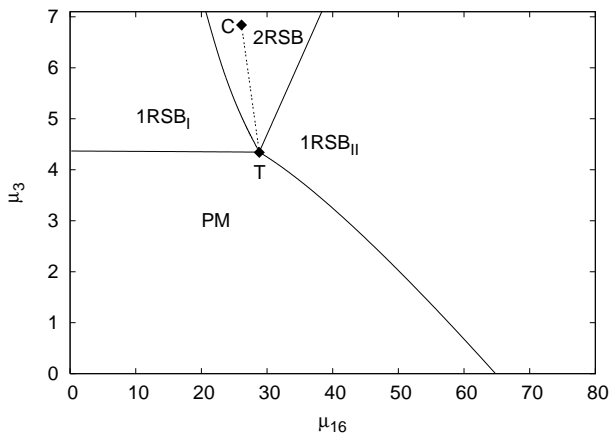}
\caption{Static phase diagram of the $3+16$ model in the ($\mu_{16}$,
$\mu_{3}$) plane in the parameter region considered by Krakoviack in
Fig. 2 of Ref. \onlinecite{comment}. 
His putative G-G transition line  (TC) is also plotted: the whole
line is contained in the  2RSB thermodynamically stable glassy phase.}
\label{fig:PhDi_sta}
\end{figure}
\begin{figure}[t!]
\includegraphics[width=.47\textwidth]{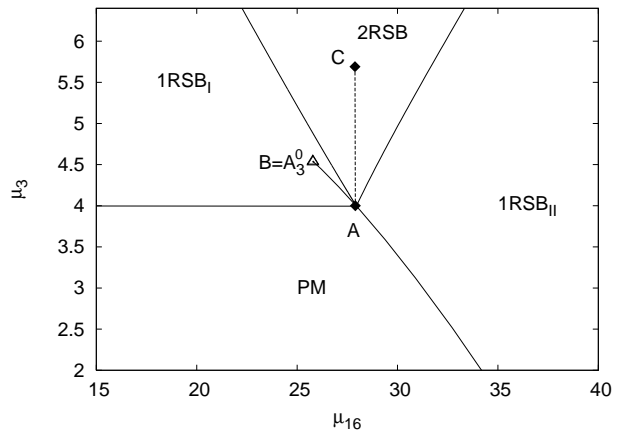}
\caption{Dynamic phase diagram of the $3+16$ model in the ($\mu_3$,
$\mu_{16}$) plane to be compared with Fig. 3 of \cite{comment}.  The
AB line corresponds to the A$_2$ transition line terminating in
B$=$A$_2^0$ that we consider in Fig. \ref{fig:3_16_a} and is not a true
transition line. The AC line is the dynamic analogue of the
TC line in the static case, cf.  Fig. \ref{fig:PhDi_sta}.}
\label{fig:PhDi_dyn}
\end{figure}

Apart from the dynamic A$_2$ line examined above,
Krakoviack\cite{comment} puts forward also other putative G-G
transition lines, both in statics and in dynamics. Without entering in
a critical analysis of their derivation, we recall that the static
line (Fig. 2 of Ref. \onlinecite{comment}) corresponds to the line
where the free energy of two different (but both 1RSB) glassy phases is equal,
that is to a first order phase transition.  Looking for RSB schemes of
computation yielding a stable thermodynamics everywhere in the
parameter space for large $p-s$, one observes, however, that a qualitatively
different phase appears in the region of the phase diagram where this
G-G transition line runs: a two step RSB {\em stable} phase.  We
reproduce Krakoviack's static G-G line in Fig. \ref{fig:PhDi_sta},
showing how it is ``eaten'' by the 2RSB phase.
\\ \indent The complete stabilized scenario for the spherical $s+p$
model (with $s,p>2$) is, actually, even more complicated than in the
$2+p$ case and leads to new results in spin-glass theory and provides
the mean-field analogues of glasses with Johari-Goldstein
processes\cite{Johari71} (i.e., thermalized processes with large
relaxation times, though shorter than the structural relaxation time),
cf. Ref. \onlinecite{unpublished}.  Here we limit ourselves to show,
in Fig. \ref{fig:PhDi_sta}, the phase diagram computed in the
framework of RSB theory around Krakoviack's putative static G-G line
and to remark that the TC line is ruled out by the presence of the
2RSB phase.
\\ 
\indent In Fig. \ref{fig:PhDi_dyn} we show the diagram for the
dynamic transition lines.  The AC line is the dynamic equivalent of
the static TC line of Fig. \ref{fig:PhDi_sta}, that is, the loci where
two apart (1RSB) solutions {\em display the same complexity}, rather
than the same free energy.  We use for the endpoints the same notation
used in Ref. \onlinecite{comment}, Fig. 3, where B is what we called
A$_3^0$ point above.  The candidate transition line AC turns out to be
embedded in the 2RSB phase (the 1RSB$_I$-2RSB and 2RSB-1RSB$_{II}$
transitions are here dynamic transitions) and is, thus, discarded.
The candidate AB line, instead, is not even a transition line, since
the complexity of {\em each} of its points is lower than the
complexity along {\em any} of the crossing (horizontal) $m$-lines.
 \\ \indent 
Although the putative G-G lines proposed by Krakoviack for
the present model and, in particular, the analogy with the G-G
transitions conjectured in the framework of MCT are ruled
out,\footnote{The disagreement with MCT must not surprise, since, in
MCT, fluid equilibrium is assumed to occur also in the frozen phase
and, hence, the equivalence between spherical $s+p$ spin glass models
and $F_{s-1,p-1}$ schematic theories only holds in the paramagnet
(fluid phase).}  we stress that this does not mean that transitions
between qualitatively different glassy phases are absent in the
spherical $s+p$ spin model with $s$ and $p$ larger than two and large
$p-s$. See, e.g., Figs. \ref{fig:PhDi_sta}-\ref{fig:PhDi_dyn}.

In summary, Krakoviack's observation that if $p-s$ is large enough G-G
transitions show up is valid. However, comparing Figs. 1, 2 and 3 of
his Comment with, respectively, Figs.  \ref{fig:3_16_a},
\ref{fig:PhDi_sta} and \ref{fig:PhDi_dyn} in this paper, one can
conclude that  those G-G transitions proposed in the Comment, and the
consequent supposed scenario, have to be rejected, even from a
heuristic point of view.

\end{document}